# THE NEAREST GROUP OF GALAXIES

Sidney van den Bergh
Dominion Astrophysical Observatory
Herzberg Institute of Astrophysics
National Research Council
5071 West Saanich Road
Victoria, British Columbia, V8X 4M6, Canada

## ABSTRACT

The small Antlia-Sextans clustering of galaxies is located at a distance of only 1.36 Mpc from the Sun, and 1.72 Mpc from the adopted barycenter of the Local Group. The latter value is significantly greater than the radius of the zero-velocity surface of the Local Group which, for an assumed age of 14 Gyr, has $R_o = 1.18 \pm 0.15$ Mpc. This, together with the observation that the members of the Ant-Sex group have a mean redshift of $+114 \pm 12$ km s$^{-1}$ relative to the centroid of the Local Group, suggests that the Antlia-Sextans group is not bound to our Local Group, and that it is expanding with the Hubble flow. If this conclusion is correct, then Antlia-Sextans may be the nearest external clustering of galaxies. The total galaxian population of the Ant-Sex group is ~ 1/5 that of the Local Group. However, the integrated luminosity of Ant-Sex is two orders of magnitude lower than that of the Local Group.

Subject headings: Galaxies - clusters: individual (Antlia-Sextans)

1.  **INTRODUCTION**

A detailed discussion of the galaxies that appear to be located along the outer fringes of the Local Group has been given by van den Bergh (1994, 2000). From these data the Local Group is found to have 32 probable members that are located within 1.0 Mpc of its barycenter. Courteau and van den Bergh (1999) find that the Local Group has a rather compact double core, which is embedded within a more extended envelope. Adopting D(M 31) = 760 kpc, and a Local Group centroid that is assumed to be located in the direction of Andromeda, at a distance of 0.6 x D(M 31), it is found that half of all Local Group members are situated within a distance of 0.45 Mpc of the adopted Local Group centroid. Since much of the mass of the Local Group is associated with M 31 at D(LG) = 0.30 Mpc, and the Galaxy at D(LG) = 0.46 Mpc, it will tentatively be assumed that the half-mass radius of the Local Group is $R_h$ ~ 0.35 Mpc. If, for the time being, we exclude the four galaxies listed in Table 1, then there are only three galaxies which are situated beyond 0.9 kpc from the Local Group centroid. They are: SagDIG with D(LG) = 1.48 Mpc, Aquarius with D(LG) = 1.02 Mpc and Tucana with D(LG) = 1.10 Mpc. These results show that the Local Group is, contrary to a widely held perception (e.g. Pritchet 1998, Jergen, Freeman & Binggeli 1999), rather well-isolated from neighboring clusters. For objects that are brighter than $M_V$ = -10.0, the number-density of all galaxies in the shell with 1.0 ≤ D(LG) < 1.5 Mpc is ~ 30 times lower than it is within the sphere with D(LG) < 1.0 Mpc.

## 2. RADIUS OF LOCAL GROUP ZERO-VELOCITY SURFACE

After excluding objects in the Antlia-Sextans group, radial velocities are presently known for 27 probable and possible Local Group members. From these objects Courteau & van den Bergh (1999) find a solar velocity of V = 306 ± 18 km s$^{-1}$ towards $\ell = 99° ± 5°$ and b = -4° ± 4°, in which the errors were estimated by using a "bootstrap" re-sampling technique. Individual Local Group members, which are plotted as filled circles in Figure 1, are found to have a radial velocity dispersion $\sigma_r$ = 61 ± 8 km s$^{-1}$ about the adopted $V_r$ versus cos θ regression line.

According to Binney & Tremaine (1987) the total mass M, the half-mass radius $R_h$, and the radial velocity dispersion are related by the relation

$$< \sigma^2 > \approx 0.4 \, G \, M \, / R_h . \qquad (1)$$

in which $<\sigma^2>$ is the mean-square speed of the system's galaxies, the components of which are assumed to be isotropic. Substituting the observational parameters given above, yields a total Local Group mass M ≈ (2.3 ± 0.6) x 10$^{12}$ M$_\odot$.

The radius $R_o$ of the zero-velocity surface, which separates the negative energy Local Group, from the surrounding positive-energy region that participates in the Hubble flow (Lynden-Bell 1981, Sandage 1986) is

$$G M T_o^2 = \pi^2 (R_o/2)^3 . \qquad (2)$$

Substituting the Local Group mass derived above this yields a radius

$$R_o (\text{Mpc}) \approx (1.18 \pm 0.15) \times [T_o (\text{Gyr})/14]^{2/3} . \qquad (3)$$

In other words the radius of the Local Group zero-velocity surface is, for an assumed age $T_o = 14$ Gyr, $R_o = 1.18 \pm 0.15$ Mpc. That the error of this value is so small is due to the fact that $R_o$ depends on the cube root of the adopted Local Group mass. The galaxies NGC 3109, Antlia, Sextans A and Sextans B, some of which had previously been regarded as possible members of the Local Group (e.g. Mateo 1998), are all found to be located well outside the zero-velocity surface of the Local Group. One should, of course, remember that the mass distribution in the Local Group is clumpy. The assumption that its zero-velocity surface is spherical is therefore an over-simplification. A more detailed discussion of the determination of the Local Group mass, and of the radius of its zero-velocity surface, is given in Courteau & van den Bergh (1999).

## 3.    THE ANTLIA - SEXTANS GROUPING

Because of their "positive velocity residuals" Yahil, Tammann & Sandage (1977) regarded NGC 3109, Sextans A and Sextans B as "Improbable Members of the Local Group". Lynden-Bell & Lin (1977) also excluded these objects from Local Group membership for the same reason. The Antlia dwarf (Whiting, Irwin & Hau 1997) is an additional probable member of this small nearby clustering. Data on each of these four cluster suspects are given below.

### 3.1 NGC 3109

This is a highly resolved galaxy, which is classified as Sm IV by Sandage & Tammann (1981). Since this object does not appear to have a nucleus van den Bergh (2000) assigns it to morphological type Ir IV. From observations of 24 Cepheids, Musella, Piotto & Capaccioli (1998) derive a distance modulus $(m-M)_o = 25.67 \pm 0.16$, corresponding to a distance of $1.36 \pm 0.10$ Mpc. This Cepheid distance is consistent with the value $(m-M)_o = 25.45$ that Richer & McCall (1992) deduced from planetary nebulae, and which Lee (1993) obtained from the magnitude of the tip of the red giant branch of NGC 3109. Adopting the Cepheid distance this galaxy has $M_V = -15.8 \pm 0.2$. Available data on NGC 3109, and other members of the Antlia - Sextans group, are summarized in Table 1.

Place Table 1 here

### 3.2 Antlia

The Antlia system is a dwarf spheroidal that exhibits no H II regions, or other signs of recent star formation. It was cataloged by Corwin, de Vaucouleurs

& de Vaucouleurs (1985), and was noted as a possible nearby galaxy by Feitzinger & Galinski (1985), and by Arp & Madore (1987). This suspicion was strengthened by Fouqué et al. (1990), who found Antlia to have a small radial velocity of $V_r = +361 \pm 3$ km s$^{-1}$. The true nature of this object was revealed by the photometric observations of individual stars by Aparicio et al. (1997), by Sarajedini, Claver & Ostheimer (1997), and by Whiting, Irwin & Hau (1997). The distance that is derived for this object depends on its adopted metallicity, and on the amount of dust that it is assumed to contain. Assuming [Fe/H] = -1.9 a distance D = 1.33 ± 0.10 Mpc is obtained.

Antlia and NGC 3109 are separated on the sky by only $1°.18$, corresponding to a projected separation of 28 kpc. Furthermore their distances (see Table 1) are, within their errors, identical. This suggests that Antlia is probably a satellite of NGC 3109. The radial velocities of NGC 3109 and Antlia differ by 43 ± 2 km s$^{-1}$. Davis et al. (1995) show, that for a gravitationally bound pair of galaxies

$$\Delta V_r^2 R_p / 2GM \leq \sin^2 \alpha \cos \alpha < 0.385, \qquad (4)$$

in which $\Delta V_r$ is the velocity difference between the galaxies, $R_p$ is their projected separation and $\alpha$ is the angle between the true line joining these two masses and the plane of the sky. Substituting the values obtained above into Eqn. (4) shows that the combined mass of NGC 3109 and Antlia must be $\gtrsim 1.6 \times 10^{10}$ $M_\odot$ for the system to be bound. The corresponding value of $M/L_V$ is $\gtrsim 90$ in solar units. This shows that the system must contain a substantial amount of dark matter for it to be gravitationally stable.

### 3.3    Sextans A

Sextans A (= DDO 75) was discovered by Zwicky (1942), who used it to argue (correctly) that the luminosity function of galaxies rises steeply towards faint absolute magnitudes, and is not Gaussian, as Hubble (1936) had claimed. The morphological classification of Sex A on the David Dunlap Observatory (DDO) system is Ir V. Sextans A was listed as a "doubtful member" of the Local Group by Humason, Mayall & Sandage (1955). From B, V, R and I photometry of a small number of Cepheids, Sakai, Madore & Freedman (1996) obtain a true distance modulus $(m-M)_o = 25.85 \pm 0.15$. Furthermore these authors find $(m-M)_o = 25.74 \pm 0.13$ from V and I photometry of the tip of the red giant branch of Sex A. The good agreement between these Population I and Population II distance indicators lends confidence in these distance determinations. A distance modulus $(m-M)_o = 25.8 \pm 0.1$, corresponding to $D = 1.45 \pm 0.07$ Mpc, will be adopted for this galaxy. Sextans A and NGC 3109 are separated by $21.\!^\circ5$, and their linear separation is ~ 500 kpc.

### 3.4 Sextans B

Sextans B (= DDO 70) is a late-type dwarf that has a DDO morphological classification Ir IV-V. From B, V, R and I photometry of Cepheids, Sakai, Madore & Freedman (1997) find a true distance modulus $(m-M)_o = 25.69 \pm 0.27$. From I photometry of the tip of its giant branch Sakai et al. derive an independent distance modulus $(m-M)_o = 25.56 \pm 0.10$ (random) $\pm 0.16$ (systematic). This value agrees, within its errors, with that obtained from Cepheids. A distance modulus of $(m-M)_o = 25.6 \pm 0.2$, corresponding to a distance of $1.32 \pm 0.12$ Mpc, has been adopted. Sex A and Sex B have similar distances, and are separated on the sky by only $10°.4$. Their linear separation is 280 kpc, and their radial velocity difference is $23 \pm 6$ km s$^{-1}$. Possibly these objects were formed close together, and subsequently drifted apart over a Hubble time at a space velocity of ~ 30 km s$^{-1}$.

## 4. CONCLUSIONS

Inspection of Figure 1 shows that the NGC 3109, Antlia, Sextans A and

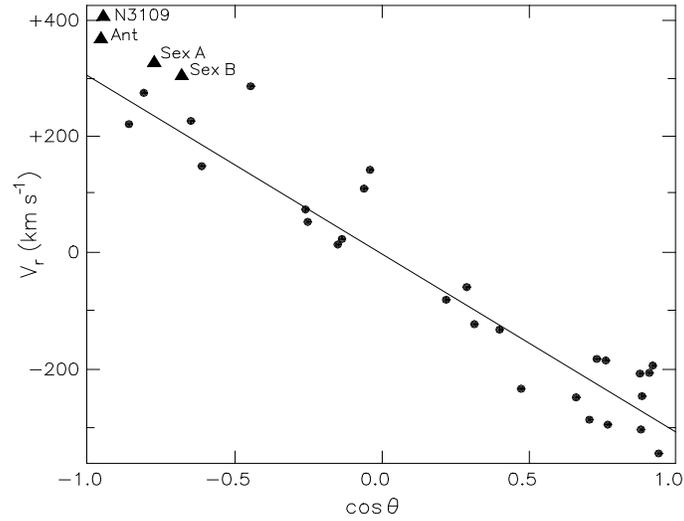

Fig. 1  Observed heliocentric radial velocity of Local Group members (filled circles), and of Antlia-Sextans galaxies (triangles) versus apex distance. The dotted lines are separated from the adopted regression relation by ± 61 km s$^{-1}$. The most deviant Local Group members are Leo I and the Sagittarius dwarf.

Sextans B form a tight clustering in the $V_r$ versus $\cos\theta$ diagram, that is located 114 ± 12 km s$^{-1}$ above the Courteau & van den Bergh (1999) relation between radial velocity and apex angle for Local Group galaxies. Taken at face, value this suggests that the Ant-Sex clustering is situated beyond the zero-velocity surface of the Local Group, and is participating in the expansion of the Universe. This is consistent with the observation that the Ant-Sex galaxies have a mean distance <D(LG)> = 1.72 Mpc from the adopted centroid of the Local Group. This is significantly larger than the 1.18 ± 0.15 Mpc radius of the Local Group zero-velocity surface derived by Courteau & van den Bergh. The D(LG) = 1.7 Mpc distance to the Ant-Sex clustering is much smaller than those to other nearby clusters, such as Sculptor (Jergen et al. 1998) [D(LG) ≈ 2.4 Mpc], IC 342/Maffei group (Krismer, Tully & Gioia 1995) [D(LG) = 3.2 Mpc] and M 81 (Freedman et al. 1994) [D(LG) = 3.4 Mpc]. This indicates that the Antlia-Sextans clustering may be the nearest neighbor of the Local Group. The number of galaxies brighter than $M_V$ = -11.0 in the Local Group is 21, compared to 4 such objects in the Antlia-Sextans group. This suggests that the total galaxian population of the Ant-Sex group is ~ 1/5 of that of the Local Group. However, because Ant-Sex does not contain giant galaxies like the Milky Way and Andromeda, its luminosity is ~ 200 times lower than that of the Local Group.

After correcting for small differences in solar apex distance, the radial velocity dispersion in Ant-Sex is found to be $\sigma_r$ = 23 km s$^{-1}$, compared to $\sigma_r$ = 61 km s$^{-1}$ for the Local Group. Since the Antlia-Sextans group has only four known members, the true values of $R_h$ and $\sigma_r$, and hence its total mass, remain extremely uncertain. Whiting (1999) has shown that it is difficult to use the deceleration of the Antlia-Sextans cluster to place meaningful constraints on the mass of the

Local Group (Sandage 1986). It would be of interest to search for additional faint members of the Ant- Sex group in Antlia, Hydra and Sextans.

I am deeply indebted to an anonymous referee for a number of corrections and very helpful suggestions, and to Stéphane Courteau for extensive discussions on the dynamics of the Local Group.

# TABLE 1

## DATA ON ANTLIA-SEXTANS MEMBERS

| Name | Alias | α (J2000) δ | Type DDO | $M_V$ | $V_r$ km s$^{-1}$ | D Mpc | D(LG) Mpc |
|---|---|---|---|---|---|---|---|
| Sextans B | DDO 70 | 09$^h$ 59$^m$ 59$^s$.9  +05° 19′ 42″ | Ir IV-V | -14.3 | +301 | 1.32 ± 0.14 | 1.60 |
| NGC 3109 | DDO 236 | 10 03 06.7  -26 09 07 | Ir V | -15.8 | +404 | 1.36 ± 0.1 | 1.75 |
| Antlia | ... | 10 03 34.3  -27 19 48 | dSph | -11.0 | +361 | 1.33 ± 0.1 | 1.72 |
| Sextans A | DDO 75 | 10 11 01.3  -04 42 48 | Ir V | -14.2 | +324 | 1.45 ± 0.07 | 1.79 |